\documentclass{article}
\usepackage{graphicx}
\usepackage[round]{natbib}   
\usepackage{subcaption}
\usepackage{amsmath,bm}
\usepackage{amssymb}

\title{Modelling size distributions of marine plastics under the influence of continuous cascading fragmentation}
\author{Mikael L. A. Kaandorp, Henk A. Dijkstra, Erik van Sebille}
\date{October 2020}

\begin{document}

\maketitle
\begin{abstract}
Field studies have shown that plastic fragments make up the majority of plastic pollution in the oceans in terms of abundance. How quickly environmental plastics fragment is not well understood, however. Here, we study this fragmentation process by considering a model which captures continuous fragmentation of particles over time in a cascading fashion. With this cascading fragmentation model we simulate particle size distributions (PSDs), specifying the abundance or mass of particles for different size classes. The fragmentation model is coupled to an environmental box model, simulating the distributions of plastic particles in the ocean, coastal waters, and on the beach. We compare the modelled PSDs to available observations, and use the results to illustrate the effect of size-selective processes such as vertical mixing in the water column and resuspension of particles from the beach into coastal waters. The model quantifies the role of fragmentation on the marine plastic mass budget: while fragmentation is a major source of (secondary) plastic particles in terms of abundance, it seems to have a minor effect on the total mass of particles larger than 0.1 mm. Future comparison to observed PSD data allow us to understand size-selective plastic transport in the environment, and potentially inform us on plastic longevity. 

\end{abstract}

\section{Introduction}

\begin{figure}
\centering
        \includegraphics[trim={2cm 0cm 3cm 0cm},clip,width=1.0\textwidth]{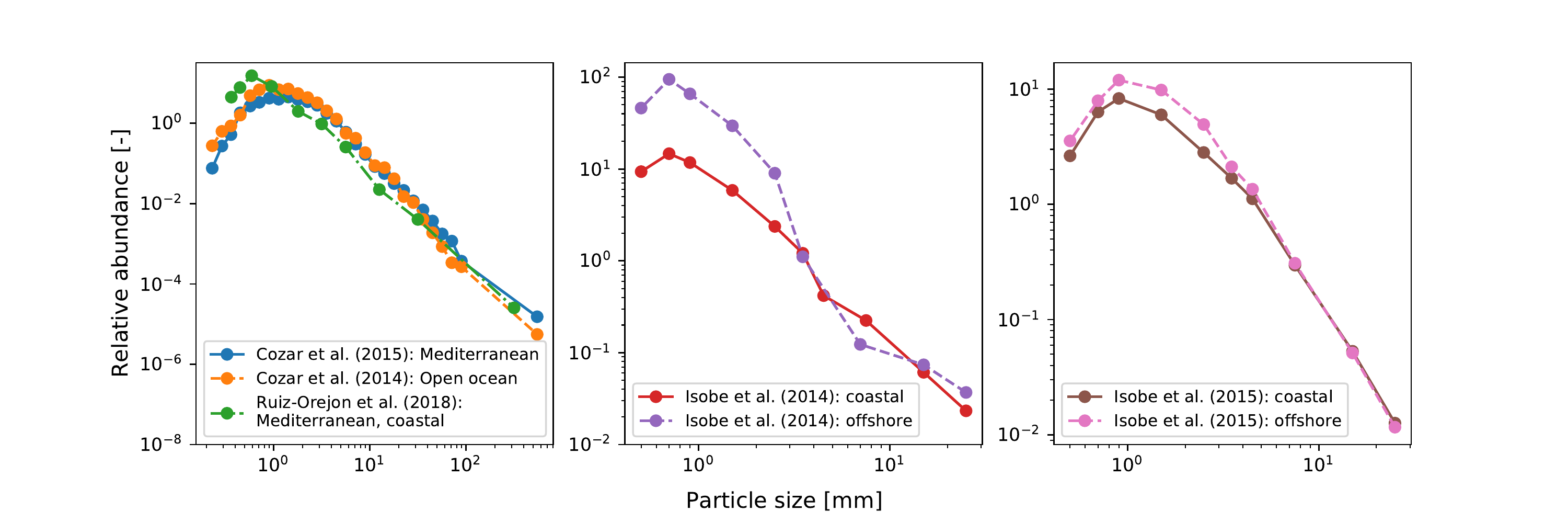}
        \caption{Observed number size distributions from \cite{Cozar2014,Cozar2015}, \cite{Ruiz-Orejon2018}, and \cite{Isobe2014,Isobe2015}. Coastal samples are defined as less than 15 kilometers from the shoreline. All particle size distributions have been normalized relative to the total abundance of large items ($>10$ mm) to show differences for small particle sizes clearly. Most datasets seem to follow a power law for the larger particle size classes (see figure~\ref{fig:results_bm_meas} below). Coastal samples tend to have relatively few small particles (or: relatively many large particles), and all distributions show a peak in abundance around 1 mm, instead of a monotonic increase with smaller sizes. }
    \label{fig:PSD_observed} 
\end{figure}

Studies have shown that fragments make up the majority of marine plastic litter in terms of abundance \citep{Cozar2015,Suaria2016}. The large amount of fragments is evident from particle size distribution (PSD) data, specifying the abundance or mass of particles for different size classes. An overview of PSD data from various studies is given in \cite{Kooi2019}; some examples are presented in figure~\ref{fig:PSD_observed}. What is commonly observed in PSD data is a power law for larger fragments ($>$1 mm in figure~\ref{fig:PSD_observed}), resulting in a straight line on a log-log scale. Oftentimes, a maximum in the PSD is observed at smaller particle sizes ($\sim$1 mm in figure~\ref{fig:PSD_observed}), ending the power law regime. 

If we want to explain the particular shapes of measured PSD data, we should first investigate the fragmentation process. Fragmentation of plastics is likely dominant on beaches, where plastics are subjected to UV-radiation, oxidation, and higher temperatures, embrittling the particles, which enhances the breaking down of particles by mechanical abrasion \citep{Andrady2011,Kalogerakis2017,Song2017,Efimova2018}. Fragmentation models have been proposed in e.g. \cite{Cozar2014}, hypothesising that the PSD slope depends on whether particles break down in a three-dimensional fashion (i.e. like a cube), or more in a two-dimensional fashion (like a thin sheet). 

While the main driver behind the PSD might be fragmentation, physical processes can have a size-selective influence on plastic particles \citep{Sebille2020}. Vertical mixing has been shown to influence PSDs measured in the water \citep{Reisser2015,Poulain2019}. Assuming similar particle shapes and densities, smaller particles tend to have lower rise velocities due to a lower buoyancy which counteracts the water friction. Turbulent mixing, induced by for example the wind, will therefore distribute these smaller particles across larger depths \citep{Kukulka2012,Chor2018}, reducing smaller size fractions in PSDs measured by nets at the ocean surface (typically submerged $\pm$10--50 centimeters depending on net type, see e.g. \cite{Pedrotti2016}, \cite{Cozar2015}, and \cite{Suaria2016}).

Size-selective lateral transport of plastic particles is likely to influence the PSD shape as well. Floating particles experience a net drift caused by waves at the ocean surface, i.e. Stokes drift \citep{Stokes1847}. Bigger (more buoyant) particles likely experience more influence from Stokes drift, given its limited depth of influence \citep{Bremer2017,Breivik2016}. Model studies have shown that Stokes drift tends to push particles towards coastal areas \citep{Iwasaki2017,Onink2019,Delandmeter2019}. In \cite{Isobe2014} it was shown that PSD data obtained close to the coast indeed showed relative overabundances of larger plastic particles. In figure~\ref{fig:PSD_observed} this is illustrated by differentiating between coastal and offshore areas: relatively high abundances of small fragments in the open ocean can be seen. 

Coastal processes, such as beaching and resuspension, can be size-selective. In \cite{Hinata2017}, residence times of particles on beaches were estimated using tagged litter. Higher particle rise velocities in the water were related to longer residence times, as these particles are more likely to be pushed to the backshore by wave swash. This could mean that larger objects remain longer on beaches, and hence experience more weathering \citep{Hinata2020}. 

Finally, PSDs could be influenced by size selective sinking, induced by for example biofouling \citep{Ryan2015}. Biofouling models predict that smaller particles, which have a larger surface to volume ratio, tend to sink more quickly \citep{Kooi2017}. This has been observed in experimental studies as well \citep{Fazey2016}.

In this work, we first consider a fragmentation model, based on fractal theory \citep{Turcotte1986,Charalambous2015}. We will use this model to simulate how PSDs resulting from fragmentation might evolve over time. Afterwards, we couple this fragmentation model to a conceptual box model, simulating movement of plastic particles between the beach, coastal water, and open ocean. Different particle sizes experience different forcings in this model, under the influence of wind-induced turbulence \citep{Kukulka2012,Poulain2019}, size-selective transport due to Stokes drift \citep{Breivik2016}, and size-selective resuspension from beaches \citep{Hinata2017}. We will compare the modelled PSDs to available observations.

Instead of counting the amount of particles in a given size-range, one can also measure their mass. In this work, we make a distinction between the two, by considering a number (i.e. abundance) size distribution (NSD), and a mass size distribution (MSD). The term PSD is used when applicable to both.

\section{Methodology}
\label{sec:methodology}

\subsection{The cascading fragmentation model}

\begin{figure}
\centering
        \includegraphics[trim={.5cm 16cm 8cm 4cm},clip,width=.8\textwidth]{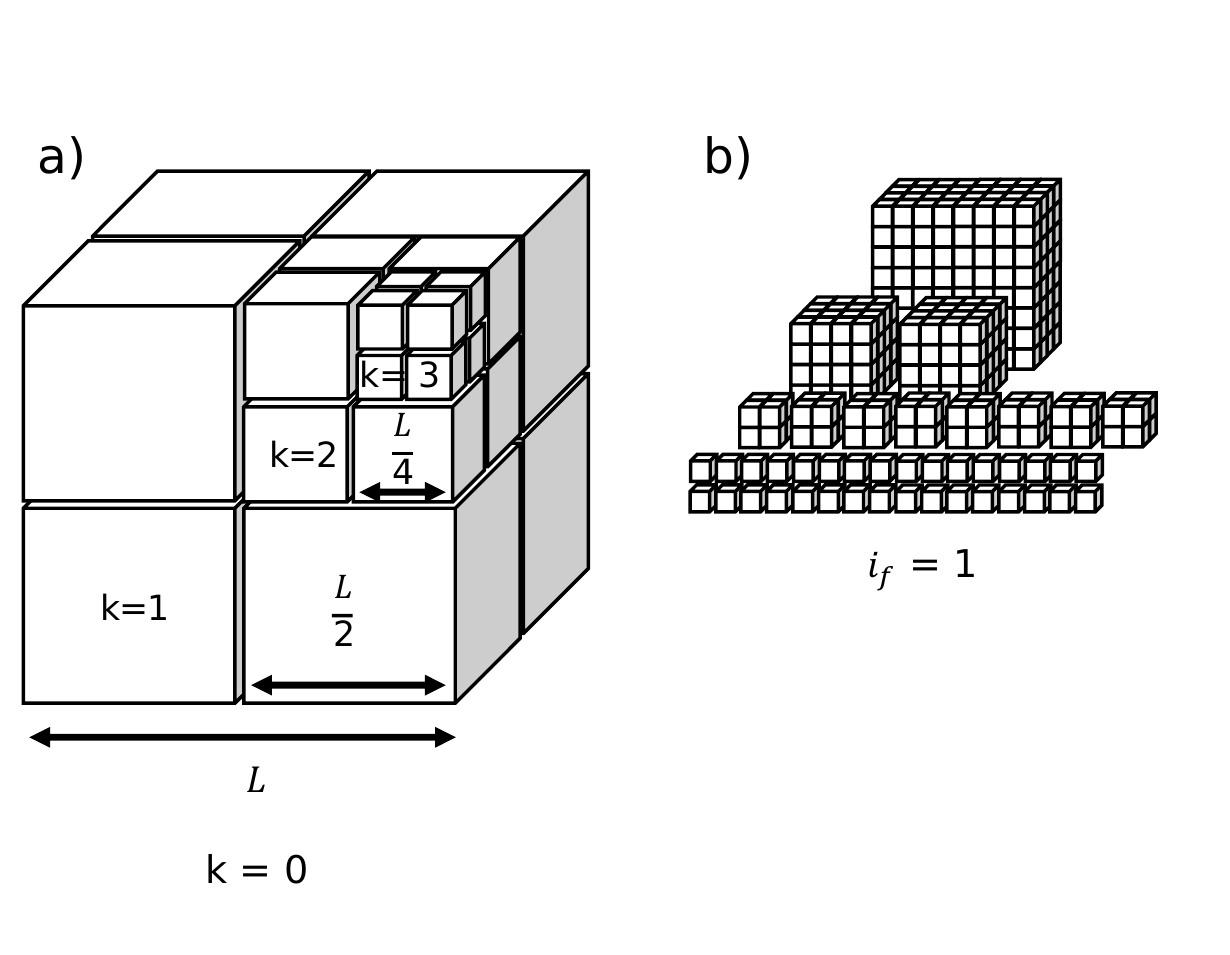}
        \caption{a) Illustration of a parent cube (size class $k=0$), consisting of successively smaller cubes, based on \cite{Turcotte1986}. Only 3 iterations until size class $k=3$ are shown here, the size class can increase indefinitely. b) Illustration of the cascading fragmentation model with $p = 0.5$ and after one fragmentation event $i_f$, based on \cite{Turcotte1986} and \cite{Charalambous2015} }
    \label{fig:fractal_cube} 
\end{figure}

The fragmentation model discussed here is based on simple fractal geometries. We define the spatial dimension as $D_N$. When $D_N = 3$, we start with a cube with a size of $L \times L \times L$ which we call the parent object, see figure~\ref{fig:fractal_cube}a. This cube can be split in eight equally-sized cubes, which can each be recursively split again. The size class of the parent object is defined as $k=0$, the size class of the cubes with length $L/2$ is defined as $k=1$, and so on. When $D_N = 2$, the starting object is a sheet instead of a cube, which can be split in four smaller sheets each time the size class increases. 

\cite{Cozar2014} presented a fragmentation model where objects are broken down into a set of smaller (equally sized) fragments in a series of successive fragmentation events. This fragmentation model was used to explain why measured PSDs often resemble power laws, i.e. functions of the form
\begin{equation}
n(l) = C l^{-\alpha},
\label{eq:powerlaw}
\end{equation}
where $n$ is the abundance, $l$ is the particle size, $\alpha$ is the power law slope, and $C$ is a constant. However, this fragmentation model requires a constant input of new parent objects to achieve a power law, while laboratory experiments have shown that power laws in the PSD also appear after fragmenting a single input of parent objects \citep{Song2017}. 

The fragmentation model used here builds upon the work of \cite{Turcotte1986}, where it was noted that scale-invariance of the fragmentation process, whether it be caused by weathering, explosions, or impacts, leads to such a power law. The idea behind the model of \cite{Turcotte1986} can be illustrated using figure~\ref{fig:fractal_cube}b. Following one fragmentation event $i_f$, a certain fraction $p$ of the original cube (size class $k=0$) splits off. For example, if $p = 0.5$, this results in 4 fragments of $k=1$ splitting off, leaving 0.5 object in size class $k=0$. This process is assumed to be scale invariant: a fraction $p$ will split off from the fragments in size class $k=1$ as well: 16 fragments of size class $k=2$ are created, and $4 \times 0.5 = 2$ fragments are left in size class $k=1$. This process is repeated indefinitely. 

\cite{Bird2009} and \cite{Gregory2012} extended this model by including a temporal component, with each fragment breaking down further as $i_f$ progresses. \cite{Charalambous2015} showed that repeatedly breaking down fragments over discrete steps of $i_f$ is a sequence of independent and identical Bernoulli trials with a chance of success $p$, yielding a negative binomial distribution. This is rewritten in terms of a continuous fragmentation index $f$ (instead of the discrete $i_f$), yielding a probability density function giving the mass $m$ in size class $k$ at fragmentation index $f$ as:
\begin{equation}
m(k; f, p) = \frac{\Gamma(k+f)}{\Gamma(k+1)\Gamma(f)} p^k (1-p)^{f},
\label{eq:pmf_m}
\end{equation}
where $\Gamma$ is the gamma function. We will call this the cascading fragmentation model. We assume that $f$ is directly proportional to time in the environment, and will review this assumption in the discussion. 

We assume $p$ to be a material property. The amount of fragments in a given size class is estimated by multiplying the mass with $2^{D_N k}$, a factor determining how many fragments of size class $k$ fit inside the parent object:
\begin{equation}
n(k, f, p) = 2^{D_N k} \, m(k; f, p)
\label{eq:pmf_n}
\end{equation}
We use $D_N = 3$ as the baseline. However, this factor is $D_N=2$ for purely flat objects like plastic sheets and $D_N=1$ for fibers or lines. As real-world samples contain a combination of these objects, the value for $D_N$ in the environment can be a non-integer between 1 and 3.

\begin{figure}
\centering
        \includegraphics[trim={2cm 0cm 1cm 0cm},clip,width=1.0\textwidth]{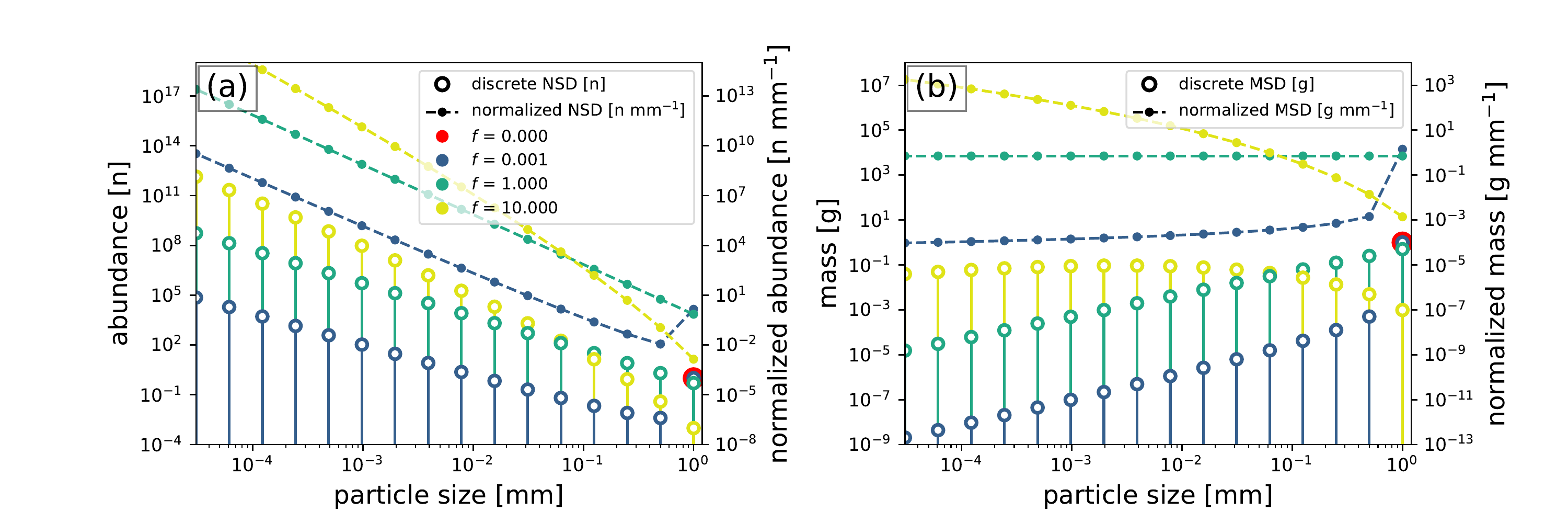}
        \caption{Particle size distributions from the cascading fragmentation model, using $p = 0.5$. Distributions for different fragmentation indices $f$ are shown. Figure a: the discrete NSD (left y-axis) and the NSD normalized with respect to the particle size (right y-axis). Figure b: the discrete MSD (left y-axis) and the normalized MSD (right y-axis) }
    \label{fig:CFM_n} 
\end{figure}

Figure~\ref{fig:CFM_n}a shows the NSD resulting from the cascading fragmentation model at various fragmentation indices $f$. We start with one cube with a length $L$ of 1 mm at $f = 0$. The continuous description in \eqref{eq:pmf_n} allows us to model the amount of fragments at a very small fragmentation index of $f = 0.001$. There are few larger fragments ($> 10^{-2}$ mm) per parent object at this stage. At $f = 1$ we have exactly a power law in the NSD, equivalent to the model by \cite{Turcotte1986}. A fractal dimension $D_f$ of the object formed by all fractions can be defined, relating to $D_N$ and $p$ by:
\begin{equation}
D_f = \mathrm{log}_2 \left( 2^{D_N} p \right).
\label{eq:D_f}
\end{equation}
The NSD power law slope at $f = 1$ is given by this fractal dimension. 

Fragments can be broken down further, eventually resulting in the NSD shown for $f = 10$. This is not a power law anymore, and the slope of this curve has increased significantly, with relatively many particles in the small size classes. The NSD (units: n) can be normalized, by dividing the amount of fragments by the size class bin width (units: n$\,$mm$^{-1}$). These normalized NSDs are presented by the dashed lines. Because of the log-scale on the x-axis, the distance between the given particle sizes increases by a constant factor. This increases the magnitude of the normalized NSD slopes by 1 compared to the discrete NSD. These normalized NSDs allow for comparison between different studies, and are therefore usually reported in literature.

Figure~\ref{fig:CFM_n}b shows the same analysis in terms of mass, i.e. the MSD, starting with one cube of 1 g and 1 mm$^3$. As fragmentation progresses, mass shifts from the large fragments towards smaller fragments. At $f = 1$ we have a power law: the difference in the slope between the NSD and MSD is 3, resulting from the $2^{D_N k}$ term in \eqref{eq:pmf_n}, with $D_N = 3$. 

\subsection{Environmental box model}
\label{sec:meth_bm}

\begin{figure}
\centering
        \includegraphics[trim={5cm 3cm 5cm 1cm},clip,width=0.6\textwidth]{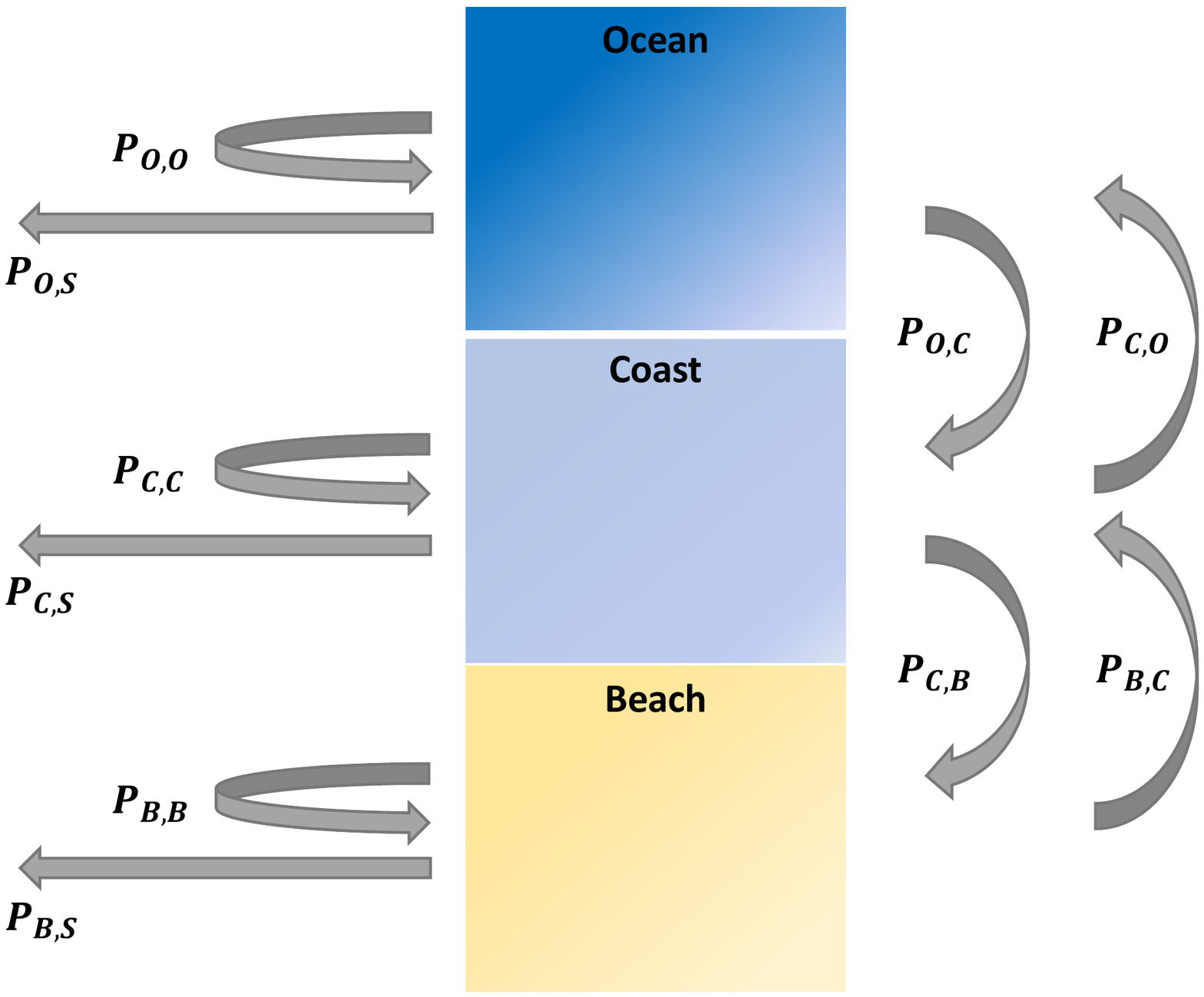}
        \caption{Environmental box model used to simulate Particle Size Distributions in the marine environment. Particles move between and within the ocean, coast, and beach boxes, and for each box there is a probability of being removed from the system.}
    \label{fig:box_model} 
\end{figure}

With the cascading fragmentation model we can now simulate PSDs over time. How particles are affected by the environment varies for different particle sizes, however. The combination of fragmentation and size-selective transport is investigated using a box model, presented in figure~\ref{fig:box_model}. The boxes in this model represent three different environmental regions: the beach, coastal waters, and open ocean. 

Particles can move between the different environments, defined by a set of transition probabilities ($P$): particles can move to a different environmental box (the arrows on the right in figure~\ref{fig:box_model}), remain in the current box (recurring arrows on the left), or vanish from the system (i.e. a sink, arrows on the left). Subscripts in figure~\ref{fig:box_model} denote ocean, coast, beach, or sink ({\it O}, {\it C}, {\it B}, and {\it S} respectively). 

Besides different environmental regions, we have different particle size classes. For a given size class, certain mass fractions will move to smaller size classes under the influence of fragmentation. These fractions are estimated by evaluating \eqref{eq:pmf_m} for the given time step of the box model. Similarly, \eqref{eq:pmf_n} is evaluated to determine the abundance of fragments moving to smaller size classes. Fragmentation is assumed to only happen on the beach, where degradation is expected to be much more effective than in the sea \citep{Efimova2018}.

Each environmental box contains a range of different particle size classes. The combination of environmental transition probabilities and fragmentation is modelled using a transition matrix. For example, taking 15 different size classes and 3 environmental compartments leads to a transition matrix of size $45 \times 45$. Further details are given in the supplementary material (section S2).

Environmental transition probabilities are based on previous findings from literature, and are discussed in the next paragraphs. A set of baseline parameters are varied in the box model to investigate their influence on the modelled PSDs. Where possible, box model parameters are calibrated for the Mediterranean sea, for which plastic sources and transport estimates are available \citep{Kaandorp2020a}. For fragmentation we have two unknown parameters: $p$ in \eqref{eq:pmf_m}, and a fragmentation rate $\lambda$ relating $f$ to time in the environment. These are estimated by comparing the model to experimental weathering data from \cite{Song2017}, presented in the results section. 

We assume that new plastic objects are introduced on the beach, and assume an initial length of 200 mm based on typical dimensions of municipal plastic waste, see the supplementary material (section S4). We use $D_N = 3$ as the baseline, i.e. cubical-shaped objects. The model time step is set to one week.

\paragraph{Ocean Transport}

A Lagrangian simulation of floating plastic in the Mediterranean, see \cite{Kaandorp2020a}, is used to determine transition probabilities within and between the ocean and coast ($P_{O,O}$, $P_{O,C}$, $P_{C,C}$, and $P_{C,O}$). The coast is defined as the ocean within 15 kilometers of the coastline. 

The baseline transition probabilities are set to a uniform value for all particle sizes, assuming they reside at the ocean surface where they experience the maximum Stokes drift. We will then evaluate the influence of size-selective lateral transport, induced by vertical mixing of particles and differences in Stokes drift over depth, see \cite{Isobe2014} and \cite{Iwasaki2017} for similar approaches. Vertical mixing of particles is estimated using the approach from \cite{Poulain2019}: for given particle sizes, the rise velocity $w_b$ is estimated by assuming ellipsoid particles with a density of $950 \, \mathrm{kg\, m^{-3}}$. From this rise velocity we estimate the median particle depth, using the particle density profiles from \cite{Kukulka2012}. The Stokes drift is estimated at this depth, assuming a Stokes profile based on the Phillips wave spectrum, see \cite{Breivik2016}. For this Stokes drift, the transition probabilities are estimated using Lagrangian model simulations \citep{Kaandorp2020a} with different Stokes drift factors. See the supplementary material (section S3) for more information.

\paragraph{Beaching and resuspension}

Using the same approach as in \cite{Kaandorp2020a}, we estimate $P_{C,B}$ by analysing drifter buoy data: from a set of 1682 drifters in the Mediterranean \citep{Menna2017}, we calculate how much time these drifters spend near the coast before beaching. For drifter buoys within 15 kilometers of the coastline, the beaching rate is estimated to be about $6.69\cdot10^{-3}\,$day$^{-1}$ (corresponding to an e-folding time-scale $\tau_{CB}$ of 149 days). In \cite{Kaandorp2020a} it was estimated that $\tau_{CB}$ for plastic particles is about 3 times lower than that for drifter buoys. We will therefore use $\tau_{CB}$ = 50 days as the baseline estimate here.

In \cite{Hinata2017}, tagged litter on beaches was tracked over 1--2 years, to estimate their residence times $\tau_{BC}$. As a baseline estimate we use $\tau_{BC} = 211$ days, reported for small plastic floats (corresponding to $P_{BC}=4.73\cdot~10^{-3}\,$day$^{-1}$). We will also look at the effect of size-selective resuspension, for which the empirical relation from \cite{Hinata2017} is used, i.e.,
\begin{equation}
\tau_{BC} (w_b) = 2.6 \cdot 10^{2}\; w_b  + 7.1,
\label{eq:Hinata}
\end{equation}
where $\tau_{BC}$ is given in days, and $w_b$ in m$\,$s$^{-1}$.

\paragraph{Sinks}

The box model also requires transition probabilities for removal of particles: $P_{O,S}$, $P_{C,S}$, $P_{B,S}$. We assume these are the same in all compartments, denoted by $P_{S}$. A given value for $P_{S}$ yields a certain amount of steady-state mass in the system. We take the estimated input of waste into the Mediterranean from \cite{Kaandorp2020a} (2,500 metric tonnes for the year 2015), and the estimated total floating mass from \cite{Cozar2015} (2,000 metric tonnes). The value for $P_{S}$ is iterated until this mass balance is satisfied, see the supplementary material (section S2) for more information.

\section{Results}

\subsection{Calibration with laboratory experiments}

In \cite{Song2017}, plastic pellets were subjected to different levels of UV exposure and to 2 months of mechanical abrasion with sand, simulating a beach environment. We use these data to calibrate the cascading fragmentation model. The data for polyethylene (PE) and polypropylene (PP) pellets (26~mm$^3$ and 19~mm$^3$ respectively) are used, as these are the most abundant polymers in surface ocean \citep{Bond2018}.

We assume a single $p$ value per material, and $D_N = 3$. The fragmentation index $f$ is allowed to vary between the different levels of UV exposure when fitting the data. By fixing $p$ and varying $f$, we get a robust estimate for the unknown parameter $p$ for which we need a plausible value in the box model. We can expect that $f$ is larger for particles subjected to longer periods of UV exposure, since embrittlement will make it easier for the mechanical abrasion to wear down the particles. 

\begin{figure}
\centering
        \includegraphics[trim={2cm .5cm 3cm 0cm},clip,width=1.0\textwidth]{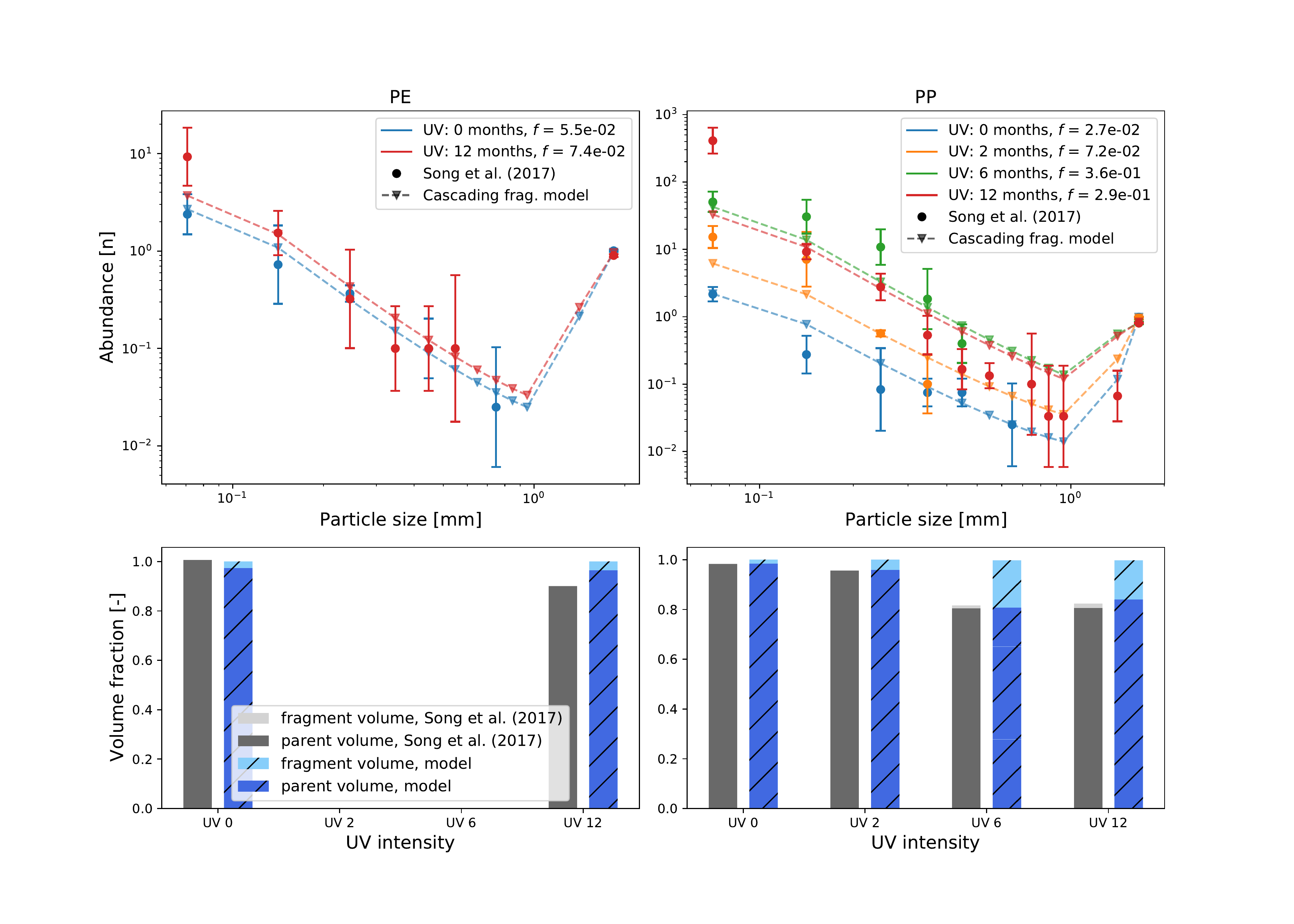}
        \caption{Calibration of the cascading fragmentation model with experimental data from \cite{Song2017}. The experimental data is shown using the circles, including errorbars representing the relative error (one time the sample standard deviation). The fragmentation model results are shown using the triangles and dashed lines.}
    \label{fig:validation_Song} 
\end{figure}

Resulting NSD fits using weighted least squares are presented in figure~\ref{fig:validation_Song}, top row, fitted values for $f$ are presented in the legend. For PE particles, the best fit results in $p$ = 0.39, for PP particles $p$ = 0.45. The experimental data are still at the early stage of fragmentation ($f < 1$), with few fragments per parent pellet, except for small fragment sizes. 

There is a good fit for the PE data, with almost all simulations within the data errorbars (one relative standard deviation). For PP there is a good fit for 0, 2 and 6 months of UV exposure. At 12 months of UV exposure there is more mismatch for the smallest size class (0.05--0.10 mm). This is also the only case where the estimated $f$ is lower than for the previous level of UV exposure. 

The bottom row of figure~\ref{fig:validation_Song} compares the estimated volume fractions of the parent pellets and the fragments. Generally, the modelled volume fraction of the parent pellet is estimated well, although there is some overprediction for PE with 12 months of UV exposure. The modelled fragment volumes are higher than the ones estimated in \cite{Song2017}. A possible explanation is that some of the larger fragments, which contribute to most of the volume when $f < 1$ (see figure~\ref{fig:CFM_n}), could have been missed in the experimental setting since there are very few of these per parent object (e.g. tenths or hundredths). 

Following this analysis, we set $p$ to 0.4 in the box model, within the range of fitted $p$ for PE and PP. A fragmentation rate $\lambda$ needs to be chosen, specifying how much $f$ increases per unit time. Following \cite{Song2017} we assume that 12 months of laboratory UV exposure is roughly related to 4.2 years of environmental exposure. Taking our estimated fragmentation indices for PE and PP results in fragmentation rates $\lambda$ of $1.8 \cdot 10^{-2}$ $f\,$year$^{-1}$ to $6.9 \cdot 10^{-2}$ $f\,$year$^{-1}$. The value for PE is used as the baseline here, given it is the most common environmental polymer \citep{Bond2018}. We acknowledge that this fragmentation rate is still very uncertain, and $f$ might vary non-linearly with time in reality.

\subsection{Modelled environmental particle size distributions}

Now that we have estimates for the transition probabilities in the box model and estimates for the fragmentation parameters, we can simulate PSDs in the environment. We will quantify the power law slope $\alpha$ of the results by numerically maximizing the log-likelihood $\ell$ of the data \citep{Virkar2014}:
\begin{equation}
\ell = n (\alpha - 1) \, \mathrm{ln} \, b_{min} + \sum_{i=\mathrm{min}}^{k} n_i \; \mathrm{ln} \left( b_i^{(1-\alpha)} - b_{i+1}^{(1-\alpha)} \right),
\label{eq:powerlaw2}
\end{equation}
where $b$ are the bin boundaries used to discretize the data, containing $n_i$ samples in the bin with index $i$, and $n = \sum n_i$. In some cases, not the entire particle size range adheres to a power law. The lower bound of the power law domain is estimated by minimizing the Kolmogorov-Smirnov statistic between the modelled NSD and the theoretical power law NSD \citep{Virkar2014}.

NSDs resulting from the box model are shown in figure~\ref{fig:results_bm_N}, corresponding MSDs and a table with parameter settings can be found in the supplementary material (section S1). 
Fragmentation is expected to increase the fraction of small particles, increasing $\alpha$ over time (see figure~\ref{fig:CFM_n}). However, environmental sinks limit the magnitude of $\alpha$: assuming a constant removal rate of plastic particles, smaller fragments, which tend to be older, have a higher probability of being removed from the environment. This combination of fragmentation and environmental sinks eventually leads to an equilibrium, or statistical steady state. This is illustrated in figure~\ref{fig:results_bm_N}a using the box model with the baseline parameters described in section~\ref{sec:meth_bm}. As time progresses, the relative proportion of fragments to parent objects increases. In this scenario, it takes on the order of years for the NSD to resemble the steady state (red dashed line). The magnitude of the environmental sinks is high enough to avoid long persistence of fragmented particles: there are still relatively many parent objects, and $\alpha = 2.57$ is still below the value derived from the fractal dimension of $\alpha = 2.67$ from \eqref{eq:D_f}. 

Steady state NSDs for different scenarios are presented in figure~\ref{fig:results_bm_N}b. Results for the baseline parameters (blue lines) almost overlap with the results where size-selective lateral transport is added to the box model, induced by vertical mixing and Stokes drift (orange lines). In the baseline scenario $\alpha = 2.57$ for all three NSDs. When adding size-selective ocean transport, larger particles tend to move more frequently from the ocean to the coast. This results in slightly more small particles in the ocean box, increasing the power law slope here to $\alpha = 2.73$. Adding size-selective resuspension of particles \citep{Hinata2017} has a strong effect (green lines). Bigger objects have longer residence times on the beach, and therefore undergo more fragmentation. This produces a large number of smaller fragments with shorter residence times, which therefore move more rapidly to the coastal and ocean cells. This near-shore trapping of larger plastic objects was already hypothesized in e.g. \cite{Isobe2014}. The empirical resuspension relation \eqref{eq:Hinata} causes the model to deviate from a power law, the domain over which $\alpha$ is calculated is shaded in green in figure~\ref{fig:results_bm_N}. The model yields $\alpha=2.69$ on the beach, which is lower than in the coastal and ocean cells (both $\alpha=3.37$). A scenario where the fragmentation rate is based on polypropylene instead of polyethylene is presented (red lines). Fragmentation breaks down the particles more quickly: a monotonic relation between particle size and abundance is observed, with $\alpha = 3.03$. Finally, a scenario is presented (purple lines) where the input of plastic waste into the Mediterranean is 100,000 tonnes per year \citep{Liubartseva2018, Jambeck2015b}, instead of the aforementioned 2,500 tonnes per year (\cite{Kaandorp2020a}). The magnitude of the sinks needs to be much larger now to attain a mass balance based on 2,000 tonnes of floating plastics \citep{Cozar2015}. Fragmentation has little time to break down the particles, resulting in relatively few fragments per parent object.

\begin{figure}
\centering
        \includegraphics[trim={1.5cm 2cm 2cm 2cm},clip,width=0.8\textwidth]{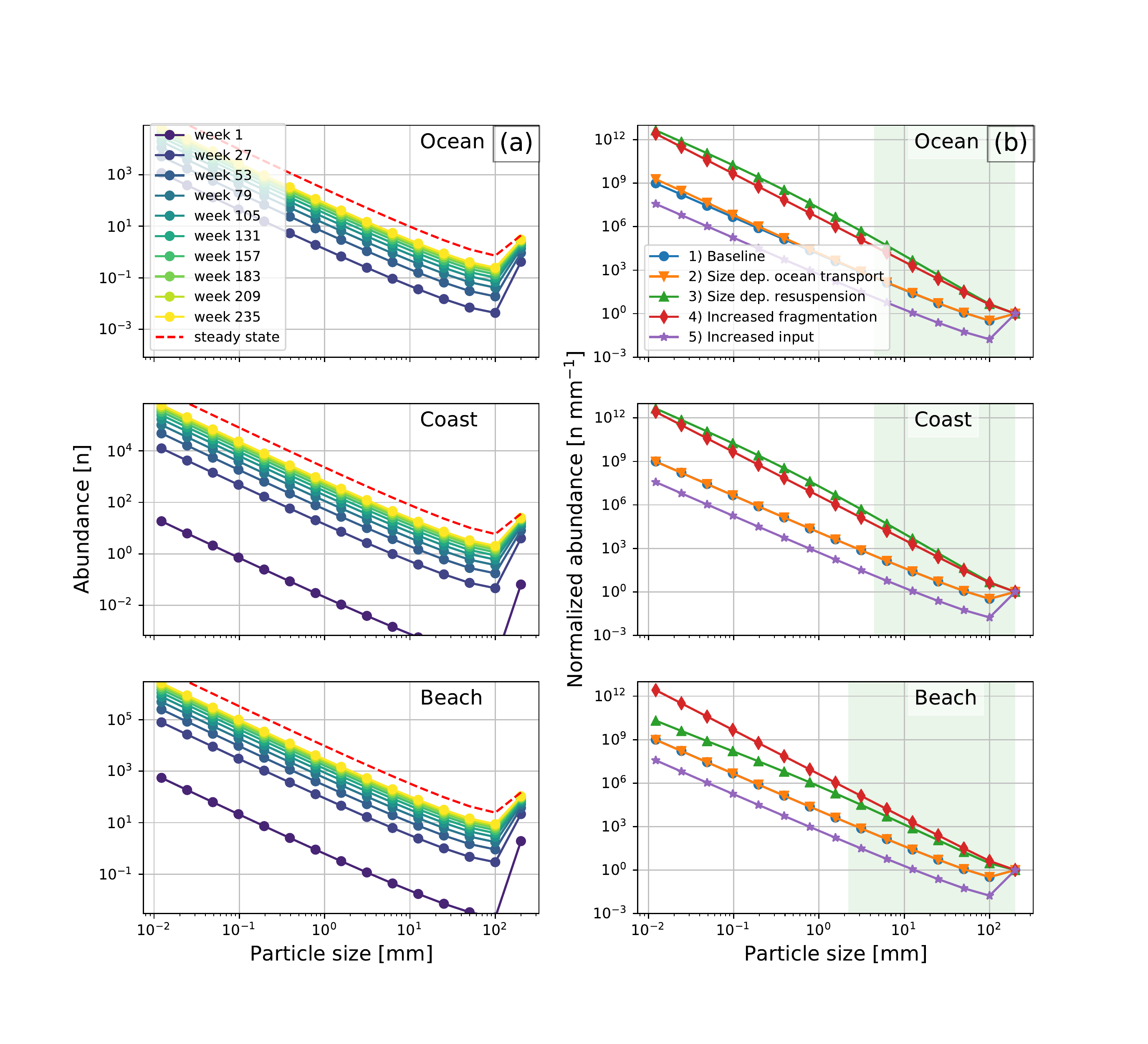}
        \caption{Modelling NSDs using the environmental box model. Column a: transient response to a constant input of particles into the model (baseline scenario). Column b: steady state normalized NSDs for different environmental scenarios, normalized to the amount of parent particles (200 mm). The approach from \cite{Virkar2014} is used to determine the power law size range: for most scenarios, $\alpha$ is calculated over the entire size range except for scenario 3, which is shown using the green shading.}
    \label{fig:results_bm_N} 
\end{figure}

\begin{figure}
\centering
        \includegraphics[trim={1.5cm 2cm 2cm 2cm},clip,width=0.8\textwidth]{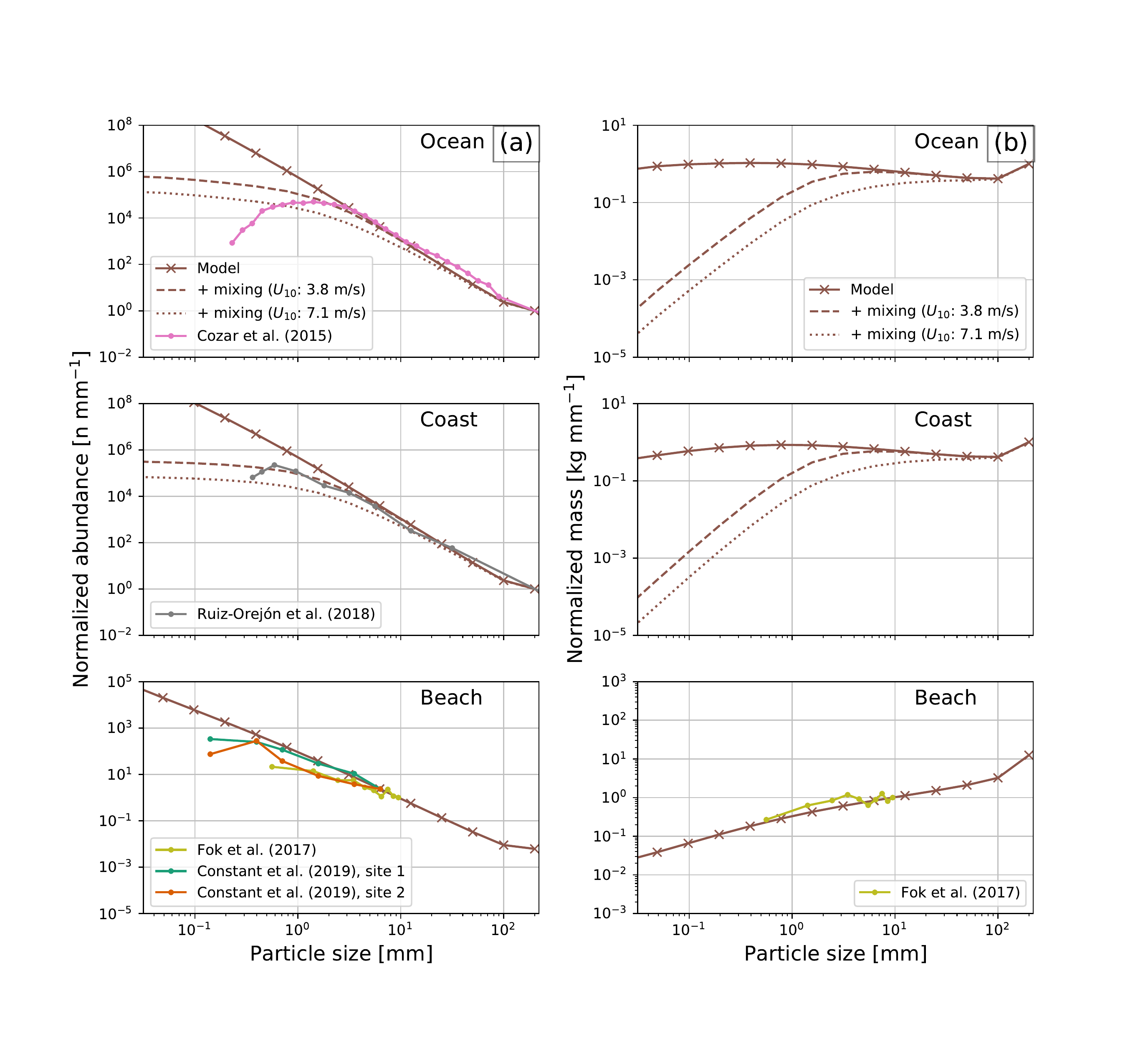}
        \caption{Comparison of modelled and measured PSDs, along with effects of vertical mixing induced by turbulence on measured particle sizes. PSDs are normalized to the maximum observed size class, if no measurements are available to the maximum modelled size class. Column a: comparison of NSDs. Column b: comparison of MSDs.}
    \label{fig:results_bm_meas} 
\end{figure}

In figure~\ref{fig:results_bm_meas}, we compare PSDs resulting from the box model with observed ones. In the model results we include both size dependent ocean transport and resuspension. Fragmentation parameters are set to $\lambda = 2\cdot10^{-4}$ $f\,$week$^{-1}$, and $D_{N} = 2.5$, resulting in good agreement with the observed PSDs. The effect of vertical turbulent mixing of fragments using the model from \cite{Poulain2019} is shown as well (calm, $U_{10} \approx 4$ m/s, and above average, $U_{10} \approx 7$ m/s conditions based on the 30\% and 70\% quantile of Mediterranean sea weather conditions \citep{Hersbach2020}, see supplementary material section S3), assuming a submerged net depth of 25 centimeters. Figure~\ref{fig:results_bm_meas}a presents the resulting NSDs. Under calm wind and wave conditions there is good agreement between the modelled and observed NSDs. Vertical mixing causes the NSD to deviate from the power law around $<$3 milimeters, similar to the NSDs measured by \cite{Cozar2015} and \cite{Ruiz-Orejon2018}. Many of the smaller fragments are expected to be mixed below the net depth, resulting in measuring only a fraction of small fragments. This, combined with a size detection limit effect where elongated particles escape from meshes smaller than their maximum length \citep{Abeynayaka2020}, could explain a part of the underabundance of sub-milimeter fragments in observations. Measurement campaigns with much smaller size-detection limits than the standard neuston nets (see e.g. \cite{Kooi2019}) show increasing abundances for sub-milimeter fragments. It is therefore unlikely that the underabundance of sub-milimeter fragments is explained by an increased loss of these particles, suggested in some studies \citep{Cozar2014,Pedrotti2016}.

Resulting power law slopes $\alpha$ are presented in table~\ref{tab:params}, as well as the estimated power law size range. The model predicts a slightly larger $\alpha$ in the ocean compared to the coast. A similar difference is seen between the chosen sets of measurements near the coast \citep{Ruiz-Orejon2018} and further away from the coast \citep{Cozar2015}, although this difference is not significant. Including vertical mixing has a strong effect on the estimated $\alpha$.

Few PSD measurements are available for beaches. Two examples are shown in figure~\ref{fig:results_bm_meas}a: one from the Mediterranean \citep{Constant2019}, and one for which both the NSD and MSD were available \citep{Fok2017}. The measurements on beaches have much lower values for $\alpha$ compared to measurements in the water. This is also captured by the model, meaning size-selective resuspension indeed seems to play an important role. At the beach the modelled $\alpha$ values are higher than the measured ones, which might indicate that size-selective beaching should be taken in account as well.

Figure~\ref{fig:results_bm_meas}b presents the modelled MSDs. Vertical mixing has a large influence on the measured mass for small particle sizes: even under calm conditions, the measured mass for particles of 0.1 mm is almost three orders of magnitude lower than without mixing. Unfortunately, there is very limited observational data reporting MSDs, so the comparison to data is more limited than for the NSDs in figure~\ref{fig:results_bm_meas}a. On beaches, the model matches the set of measurement well, but more data are necessary to further verify this. Large fragments are expected to dominate in terms of mass on beaches. In the water, $\alpha$ seems to be approximately zero on average. This would mean that the mass contribution would scale roughly quadratically for an increasing size class $k$, i.e. large fragments also dominate in terms of mass here. 

\begin{table}
\caption{Power law slopes $\alpha$ obtained by maximizing \eqref{eq:powerlaw2} to the modelled and measured PSDs. The size range for which the power law holds is estimated by minimizing the Kolmogorov-Smirnov statistic of the observed data with respect to the fitted power law, see \cite{Virkar2014}. For beach samples no lower bound is estimated given the low amount of reported bins. Uncertainties in $\alpha$ ($\pm$ 1 times the standard deviation) are not calculated for the model, since this depends on the amount of samples in the bins, which is undefined in this case.}
\centering
\label{tab:params}
    \begin{tabular}{l|p{2cm}|p{2cm}}
    \textbf{Domain} & $\bm\alpha$ & \textbf{Minimum size} \\ \hline
	Ocean & 2.73 & $\geq 3.13$ mm\\
	Ocean + mixing ($U_{10} \approx 4$ m/s) & 2.63 & $\geq 6.25$ mm\\
	Ocean + mixing ($U_{10} \approx 7$ m/s) & 2.37 & $\geq 12.5$ mm\\
	\cite{Cozar2015}  & $2.53 \pm 0.04$ & $\geq 3.16$ mm \\ \hline
	Coast & 2.69 & $\geq 3.13$ mm\\
	Coast + mixing ($U_{10} \approx 4$ m/s) & 2.60 & $\geq 6.25$ mm\\
	Coast + mixing ($U_{10} \approx 7$ m/s) & 2.34 & $\geq 12.5$ mm\\
	\cite{Ruiz-Orejon2018} & $2.49 \pm 0.06$ & $\geq 2.50$ mm \\ \hline
	Beach & 2.02 & $\geq 1.56$ mm \\
	\cite{Fok2017} & $1.60 \pm 0.01$ & $\geq 0.32$ mm \\
	\cite{Constant2019} & $1.47\pm 0.01$, $1.45\pm 0.03$ & $\geq 0.06$ mm
    \end{tabular}
\end{table}

The environmental box model used to model the PSDs is a useful tool for future mass balance studies. The steady-state with the model settings used for figure~\ref{fig:results_bm_meas}, gives that about 98\% of the mass in the system is on the beach, about 2\% in the coastal surface water, and about 0.2\% in the surface open ocean. This large fraction of plastics stranding is in good agreement with previous mass balance estimates \citep{Lebreton2019}. It should be noted that other environmental regions, like the ocean floor, are not included in these numbers as these are part of the sinks in the box model ($P_S$), which continuously take up more mass over time. 
Secondary microplastics generation can be estimated: for the same model settings, about $6.5 \cdot 10^{-5}\%$ of the macroplastic ($>5$ mm) mass breaks down into microplastics per week, about $2.0 \cdot 10^{-6}\%$ of microplastics become smaller than 0.1 mm. This is orders of magnitude smaller than the estimated sinks, taking up about $5.0 \cdot 10^{-2}\%$ of the plastic mass per week. Longevity of plastics can be estimated: taking a sudden stop of new plastics, it would take about 176 years for 99\% of the plastic mass to disappear from the surface water and beaches. This is a much longer time scale than given by the conceptual model in \cite{Koelmans2017}, where for a similar stop of new plastics, almost all plastic mass was removed from the ocean surface layer within three years. 

\section{Conclusions and discussion}

\title{Modelling size distributions of marine plastics under the influence of continuous cascading fragmentation}rtical mixing in the water column has a strong impact on measured PSDs. 

Understanding the nature of PSDs and how they differ in environmental regions can help close the marine plastic mass budget. It can be estimated which particle sizes contribute to most of the mass, as well as how persistent these are. The environmental box model is highly idealized using e.g. the assumption that the removal rate is the same in all compartments. Combined with more data and complex models \citep{Kaandorp2020a}, this type of box model can be very valuable because different input and fragmentation scenarios can quickly be evaluated to assess whether these lead to realistic PSDs. 

The environmental box model was calibrated to Mediterranean sea data as much as possible. However, resuspension probabilities from \cite{Hinata2017} were estimated on beaches in Japan, and in \cite{Song2017} the experimental weathering was related to environmental time in South Korea. Further studies are necessary on how the parameters might vary globally. 

The fragmentation model introduced in \cite{Cozar2014} focused mainly on spatial dimensionality: $\alpha=3$ in the NSD was related to three-dimensional fragmentation, i.e. a cube splitting into 8 smaller cubes. Care should be taken in future studies that when working with logarithmic binning, the normalized NSD (units: n$\,$mm$^{-1}$) slope decreases by one compared to the discrete NSD (units: n), see figure~\ref{fig:CFM_n}. This was overlooked in \cite{Cozar2014}: $\alpha = 3$ would correspond to two-dimensional fragmentation with their model, see the supplementary material (section S5) for further explanation. Normalization is also important to take in account when describing plastic particle size in terms of a probability density function \citep{Kooi2019}, specifying the probability per unit length (units: mm$^{-1}$). Finally, estimating $\alpha$ is not trivial: fitting straight lines on log-log transformed data induces large biases, maximum likelihood approaches are more suitable, see e.g. \cite{Newman2005} and \cite{Virkar2014}. 

The cascading fragmentation model by \cite{Charalambous2015} used in this work, is able to model many small chips breaking away from a parent object, and showed good correspondence with experimental data from \cite{Song2017} (figure~\ref{fig:validation_Song}). In the cascading fragmentation model, the fraction of small particles keeps increasing over time, increasing $\alpha$. We hypothesise that the value of $\alpha$ in the environment remains limited, due to the fact that smaller particles tend to be older, and hence are more likely to be removed by environmental sinks. 

Future studies should further validate the cascading fragmentation model. We assumed that $f$ and time are directly proportional here, which does not necessarily needs to be the case. In e.g. \cite{Charalambous2015}, it was shown that grinding can become less efficient as particles become smaller, which might lead to a logarithmic relation instead. Fragmentation of plastics is interesting in this sense: oxidation of plastics causes a brittle surface layer with microcracks \citep{Andrady2011}, perhaps increasing the fragmentation rate below a certain length scale, dependent on how far UV radiation penetrates the polymer. 

MSD data is currently lacking: more data would help us to verify whether the larger size classes indeed make up most of the environmental plastic mass. More PSD data from beaches would allow for better constraining residence times of plastic particles on beaches and in coastal waters, and more data from marine sediment might give insight in the role of size-selective sinking, induced by e.g. biofouling \citep{Kooi2017}.

\subsection*{Acknowledgements}
This work was supported through funding from the European Research Council (ERC) under the European Union Horizon 2020 research and innovation programme (grant agreement No 715386). This work was carried out on the Dutch national e-infrastructure with the support of SURF Cooperative (project no. 16371). We would like to thank Dr. Wonjoon Shim and Dr. Young Kyoung Song for their data on plastic fragmentation, Marie Poulain for providing data on particle sizes and vertical mixing, Dr. Andrés Cózar and Dr. Atsuhiko Isobe for providing raw particle size distribution data.

\bibliographystyle{plainnat}
\bibliography{library}

\begin{thebibliography}{45}
\providecommand{\natexlab}[1]{#1}
\providecommand{\url}[1]{\texttt{#1}}
\expandafter\ifx\csname urlstyle\endcsname\relax
  \providecommand{\doi}[1]{doi: #1}\else
  \providecommand{\doi}{doi: \begingroup \urlstyle{rm}\Url}\fi

\bibitem[Abeynayaka et~al.(2020)Abeynayaka, Kojima, Miwa, Ito, Nihei, Fukunaga,
  Yashima, and Itsubo]{Abeynayaka2020}
Amila Abeynayaka, Fujio Kojima, Yoshikazu Miwa, Nobuhiro Ito, Yasuo Nihei,
  Yu~Fukunaga, Yuga Yashima, and Norihiro Itsubo.
\newblock {Rapid sampling of suspended and floating microplastics in
  challenging riverine and coastal water environments in Japan}.
\newblock \emph{Water}, 12\penalty0 (1903), 2020.
\newblock ISSN 20734441.
\newblock \doi{10.3390/w12071903}.

\bibitem[Andrady(2011)]{Andrady2011}
Anthony~L. Andrady.
\newblock {Microplastics in the marine environment}.
\newblock \emph{Marine Pollution Bulletin}, 62\penalty0 (8):\penalty0
  1596--1605, 2011.
\newblock ISSN 0025326X.
\newblock \doi{10.1016/j.marpolbul.2011.05.030}.
\newblock URL \url{http://dx.doi.org/10.1016/j.marpolbul.2011.05.030}.

\bibitem[Bird et~al.(2009)Bird, Tarquis, and Whitmore]{Bird2009}
N~R~A Bird, A~M Tarquis, and A~P Whitmore.
\newblock {Modeling Dynamic Fragmentation of Soil}.
\newblock \emph{Vadose Zone Journal}, 8\penalty0 (1):\penalty0 197--201, 2009.
\newblock \doi{10.2136/vzj2008.0046}.

\bibitem[Bond et~al.(2018)Bond, Ferrandiz-mas, Felipe-sotelo, and
  Sebille]{Bond2018}
Tom Bond, Veronica Ferrandiz-mas, M{\'{o}}nica Felipe-sotelo, and Erik~Van
  Sebille.
\newblock {The occurrence and degradation of aquatic plastic litter based on
  polymer physicochemical properties : A review}.
\newblock \emph{Critical Reviews in Environmental Science and Technology},
  48\penalty0 (7-9):\penalty0 685--722, 2018.
\newblock \doi{10.1080/10643389.2018.1483155}.

\bibitem[Breivik et~al.(2016)Breivik, Bidlot, and Janssen]{Breivik2016}
{\O}yvind Breivik, Jean~Raymond Bidlot, and Peter~A.E.M. Janssen.
\newblock {A Stokes drift approximation based on the Phillips spectrum}.
\newblock \emph{Ocean Modelling}, 100:\penalty0 49--56, 2016.
\newblock ISSN 14635003.
\newblock \doi{10.1016/j.ocemod.2016.01.005}.

\bibitem[Bremer and Breivik(2017)]{Bremer2017}
T~S Van~Den Bremer and {\O}~Breivik.
\newblock {Stokes drift Subject Areas :}.
\newblock \emph{Philosophical Transactions A}, 376\penalty0 (20170104), 2017.

\bibitem[Charalambous(2015)]{Charalambous2015}
Constantinos Charalambous.
\newblock \emph{{On the Evolution of Particle Fragmentation With Applications
  to Planetary Surfaces}}.
\newblock PhD thesis, Imperial College London, 2015.

\bibitem[Chor et~al.(2018)Chor, Yang, Meneveau, and Chamecki]{Chor2018}
Tomas Chor, Di~Yang, Charles Meneveau, and Marcelo Chamecki.
\newblock {A Turbulence Velocity Scale for Predicting the Fate of Buoyant
  Materials in the Oceanic Mixed Layer}.
\newblock \emph{Geophysical Research Letters}, 45, 2018.
\newblock ISSN 19448007.
\newblock \doi{10.1029/2018GL080296}.

\bibitem[Constant et~al.(2019)Constant, Kerherv{\'{e}},
  Mino-Vercellio-Verollet, Dumontier, {S{\`{a}}nchez Vidal}, Canals, and
  Heussner]{Constant2019}
Mel Constant, Philippe Kerherv{\'{e}}, Morgan Mino-Vercellio-Verollet, Marc
  Dumontier, Anna {S{\`{a}}nchez Vidal}, Miquel Canals, and Serge Heussner.
\newblock {Beached microplastics in the Northwestern Mediterranean Sea}.
\newblock \emph{Marine Pollution Bulletin}, 142\penalty0 (March):\penalty0
  263--273, 2019.
\newblock ISSN 18793363.
\newblock \doi{10.1016/j.marpolbul.2019.03.032}.
\newblock URL \url{https://doi.org/10.1016/j.marpolbul.2019.03.032}.

\bibitem[C{\'{o}}zar et~al.(2014)C{\'{o}}zar, Echevarr{\'{i}}a,
  Gonz{\'{a}}lez-Gordillo, Irigoien, Ubeda, Hern{\'{a}}ndez-Le{\'{o}}n, Palma,
  Navarro, Garc{\'{i}}a-de Lomas, Ruiz, Fern{\'{a}}ndez-de Puelles, and
  Duarte]{Cozar2014}
Andr{\'{e}}s C{\'{o}}zar, Fidel Echevarr{\'{i}}a, J~Ignacio
  Gonz{\'{a}}lez-Gordillo, Xabier Irigoien, B{\'{a}}rbara Ubeda, Santiago
  Hern{\'{a}}ndez-Le{\'{o}}n, Alvaro~T Palma, Sandra Navarro, Juan
  Garc{\'{i}}a-de Lomas, Andrea Ruiz, Mar{\'{i}}a~L Fern{\'{a}}ndez-de Puelles,
  and Carlos~M Duarte.
\newblock {Plastic debris in the open ocean.}
\newblock \emph{Proceedings of the National Academy of Sciences of the United
  States of America}, 111\penalty0 (28):\penalty0 10239--44, 2014.
\newblock ISSN 1091-6490.
\newblock \doi{10.1073/pnas.1314705111}.
\newblock URL
  \url{http://www.ncbi.nlm.nih.gov/pubmed/24982135{\%}0Ahttp://www.pubmedcentral.nih.gov/articlerender.fcgi?artid=PMC4104848}.

\bibitem[C{\'{o}}zar et~al.(2015)C{\'{o}}zar, Sanz-Mart{\'{i}}n, Mart{\'{i}},
  Gonz{\'{a}}lez-Gordillo, Ubeda, {\'{A}}.g{\'{a}}lvez, Irigoien, and
  Duarte]{Cozar2015}
Andr{\'{e}}s C{\'{o}}zar, Marina Sanz-Mart{\'{i}}n, Elisa Mart{\'{i}},
  J.~Ignacio Gonz{\'{a}}lez-Gordillo, B{\'{a}}rbara Ubeda, Jos{\'{e}}
  {\'{A}}.g{\'{a}}lvez, Xabier Irigoien, and Carlos~M. Duarte.
\newblock {Plastic accumulation in the mediterranean sea}.
\newblock \emph{PLoS ONE}, 10\penalty0 (4):\penalty0 1--12, 2015.
\newblock ISSN 19326203.
\newblock \doi{10.1371/journal.pone.0121762}.

\bibitem[Delandmeter and Sebille(2019)]{Delandmeter2019}
Philippe Delandmeter and Erik~Van Sebille.
\newblock {The Parcels v2 . 0 Lagrangian framework : new field interpolation
  schemes}.
\newblock \emph{Geoscientific Model Development}, 12:\penalty0 3571--3584,
  2019.
\newblock \doi{doi.org/10.5194/gmd-12-3571-2019}.

\bibitem[Efimova et~al.(2018)Efimova, Bagaeva, Bagaev, Kileso, and
  Chubarenko]{Efimova2018}
Irina Efimova, Margarita Bagaeva, Andrei Bagaev, Alexander Kileso, and Irina~P
  Chubarenko.
\newblock {Secondary Microplastics Generation in the Sea Swash Zone With Coarse
  Bottom Sediments : Laboratory Experiments}.
\newblock \emph{Frontiers in Marine Science}, 5\penalty0 (313), 2018.
\newblock \doi{10.3389/fmars.2018.00313}.

\bibitem[Fazey and Ryan(2016)]{Fazey2016}
Francesca~M.C. Fazey and Peter~G. Ryan.
\newblock {Biofouling on buoyant marine plastics: An experimental study into
  the effect of size on surface longevity}.
\newblock \emph{Environmental Pollution}, 210:\penalty0 354--360, 2016.
\newblock ISSN 18736424.
\newblock \doi{10.1016/j.envpol.2016.01.026}.
\newblock URL \url{http://dx.doi.org/10.1016/j.envpol.2016.01.026}.

\bibitem[Fok et~al.(2017)Fok, Cheung, Tang, and Li]{Fok2017}
Lincoln Fok, Pui~Kwan Cheung, Guangda Tang, and Wai~Chin Li.
\newblock {Size distribution of stranded small plastic debris on the coast of
  Guangdong, South China}.
\newblock \emph{Environmental Pollution}, 220:\penalty0 407--412, 2017.
\newblock ISSN 18736424.
\newblock \doi{10.1016/j.envpol.2016.09.079}.

\bibitem[Gregory et~al.(2012)Gregory, Bird, Watts, and Whitmore]{Gregory2012}
A~S Gregory, N~R~A Bird, C~W Watts, and A~P Whitmore.
\newblock {An assessment of a new model of dynamic fragmentation of soil with
  test data}.
\newblock \emph{Soil and Tillage Research}, 120\penalty0 (April):\penalty0
  61--68, 2012.
\newblock \doi{10.2136/vzj2008.0046}.

\bibitem[Hersbach et~al.(2020)Hersbach, Bell, Berrisford, Hirahara,
  Hor{\'{a}}nyi, Mu{\~{n}}oz‐Sabater, Nicolas, Peubey, Radu, Schepers,
  Simmons, Soci, Abdalla, Abellan, Balsamo, Bechtold, Biavati, Bidlot,
  Bonavita, Chiara, Dahlgren, Dee, Diamantakis, Dragani, Flemming, Forbes,
  Fuentes, Geer, Haimberger, Healy, Hogan, H{\'{o}}lm, Janiskov{\'{a}}, Keeley,
  Laloyaux, Lopez, Lupu, Radnoti, Rosnay, Rozum, Vamborg, Villaume, and
  Th{\'{e}}paut]{Hersbach2020}
Hans Hersbach, Bill Bell, Paul Berrisford, Shoji Hirahara, Andr{\'{a}}s
  Hor{\'{a}}nyi, Joaqu{\'{i}}n Mu{\~{n}}oz‐Sabater, Julien Nicolas, Carole
  Peubey, Raluca Radu, Dinand Schepers, Adrian Simmons, Cornel Soci, Saleh
  Abdalla, Xavier Abellan, Gianpaolo Balsamo, Peter Bechtold, Gionata Biavati,
  Jean Bidlot, Massimo Bonavita, Giovanna Chiara, Per Dahlgren, Dick Dee,
  Michail Diamantakis, Rossana Dragani, Johannes Flemming, Richard Forbes,
  Manuel Fuentes, Alan Geer, Leo Haimberger, Sean Healy, Robin~J. Hogan,
  El{\'{i}}as H{\'{o}}lm, Marta Janiskov{\'{a}}, Sarah Keeley, Patrick
  Laloyaux, Philippe Lopez, Cristina Lupu, Gabor Radnoti, Patricia Rosnay,
  Iryna Rozum, Freja Vamborg, Sebastien Villaume, and Jean‐No{\"{e}}l
  Th{\'{e}}paut.
\newblock {The ERA5 global reanalysis}.
\newblock \emph{Quarterly Journal of the Royal Meteorological Society},
  146:\penalty0 1999--2049, jul 2020.
\newblock ISSN 0035-9009.
\newblock \doi{10.1002/qj.3803}.
\newblock URL \url{https://onlinelibrary.wiley.com/doi/abs/10.1002/qj.3803}.

\bibitem[Hinata et~al.(2017)Hinata, Mori, Ohno, Miyao, and Kataoka]{Hinata2017}
Hirofumi Hinata, Keita Mori, Kazuki Ohno, Yasuyuki Miyao, and Tomoya Kataoka.
\newblock {An estimation of the average residence times and onshore-offshore
  diffusivities of beached microplastics based on the population decay of
  tagged meso- and macrolitter}.
\newblock \emph{Marine Pollution Bulletin}, 122\penalty0 (1-2):\penalty0
  17--26, 2017.
\newblock ISSN 0025-326X.
\newblock \doi{10.1016/j.marpolbul.2017.05.012}.
\newblock URL \url{http://dx.doi.org/10.1016/j.marpolbul.2017.05.012}.

\bibitem[Hinata et~al.(2020)Hinata, Sagawa, Kataoka, and Takeoka]{Hinata2020}
Hirofumi Hinata, Nao Sagawa, Tomoya Kataoka, and Hidetaka Takeoka.
\newblock {Numerical modeling of the beach process of marine plastics : A
  probabilistic and diagnostic approach with a particle tracking method}.
\newblock \emph{Marine Pollution Bulletin}, 152\penalty0 (October
  2019):\penalty0 110910, 2020.
\newblock ISSN 0025-326X.
\newblock \doi{10.1016/j.marpolbul.2020.110910}.
\newblock URL \url{https://doi.org/10.1016/j.marpolbul.2020.110910}.

\bibitem[Isobe et~al.(2014)Isobe, Kubo, Tamura, Nakashima, and
  Fujii]{Isobe2014}
Atsuhiko Isobe, Kenta Kubo, Yuka Tamura, Etsuko Nakashima, and Naoki Fujii.
\newblock {Selective transport of microplastics and mesoplastics by drifting in
  coastal waters}.
\newblock \emph{Marine Pollution Bulletin}, 89\penalty0 (1-2):\penalty0
  324--330, 2014.
\newblock ISSN 0025-326X.
\newblock \doi{10.1016/j.marpolbul.2014.09.041}.
\newblock URL \url{http://dx.doi.org/10.1016/j.marpolbul.2014.09.041}.

\bibitem[Isobe et~al.(2015)Isobe, Uchida, Tokai, and Iwasaki]{Isobe2015}
Atsuhiko Isobe, Keiichi Uchida, Tadashi Tokai, and Shinsuke Iwasaki.
\newblock {East Asian seas : A hot spot of pelagic microplastics East Asian
  seas : A hot spot of pelagic microplastics}.
\newblock \emph{Marine Pollution Bulletin}, 101\penalty0 (2):\penalty0
  618--623, 2015.
\newblock ISSN 0025-326X.
\newblock \doi{10.1016/j.marpolbul.2015.10.042}.
\newblock URL \url{http://dx.doi.org/10.1016/j.marpolbul.2015.10.042}.

\bibitem[Iwasaki et~al.(2017)Iwasaki, Isobe, Kako, Uchida, and
  Tokai]{Iwasaki2017}
Shinsuke Iwasaki, Atsuhiko Isobe, Shin'ichiro Kako, Keiichi Uchida, and Tadashi
  Tokai.
\newblock {Fate of microplastics and mesoplastics carried by surface currents
  and wind waves: A numerical model approach in the Sea of Japan}.
\newblock \emph{Marine Pollution Bulletin}, 121\penalty0 (1-2):\penalty0
  85--96, 2017.
\newblock ISSN 18793363.
\newblock \doi{10.1016/j.marpolbul.2017.05.057}.

\bibitem[Jambeck et~al.(2015)Jambeck, Geyer, Wilcox, Siegler, Perryman,
  Andrady, Narayan, and Law]{Jambeck2015b}
Jenna~R. Jambeck, Roland Geyer, Chris Wilcox, Theodore~R. Siegler, Miriam
  Perryman, Anthony Andrady, Ramani Narayan, and Kara~Lavender Law.
\newblock {Plastic waste inputs from land into the ocean}.
\newblock \emph{Science}, 347\penalty0 (6223):\penalty0 768--771, 2015.
\newblock ISSN 10959203.
\newblock \doi{10.1126/science.1260352}.

\bibitem[Kaandorp et~al.(2020)Kaandorp, Dijkstra, and van
  Sebille]{Kaandorp2020a}
Mikael L~A Kaandorp, Henk~A Dijkstra, and Erik van Sebille.
\newblock {Closing the Mediterranean Marine Floating Plastic Mass Budget :
  Inverse Modelling of Sources and Sinks}.
\newblock \emph{Preprint}, 2020.

\bibitem[Kalogerakis et~al.(2017)Kalogerakis, Karkanorachaki, Kalogerakis,
  Triantafyllidi, Gotsis, Partsinevelos, and Fava]{Kalogerakis2017}
Nicolas Kalogerakis, Katerina Karkanorachaki, G~Calypso Kalogerakis, Elisavet~I
  Triantafyllidi, Alexandros~D Gotsis, Panagiotis Partsinevelos, and Fabio
  Fava.
\newblock {Microplastics Generation : Onset of Fragmentation of Polyethylene
  Films in Marine Environment Mesocosms}.
\newblock \emph{Frontiers in Marine Science}, 4\penalty0 (84), 2017.
\newblock \doi{10.3389/fmars.2017.00084}.

\bibitem[Koelmans et~al.(2017)Koelmans, Kooi, Law, and {Van
  Sebille}]{Koelmans2017}
Albert~A. Koelmans, Merel Kooi, Kara~Lavender Law, and Erik {Van Sebille}.
\newblock {All is not lost: Deriving a top-down mass budget of plastic at sea}.
\newblock \emph{Environmental Research Letters}, 12\penalty0 (11), 2017.
\newblock ISSN 17489326.
\newblock \doi{10.1088/1748-9326/aa9500}.

\bibitem[Kooi and Koelmans(2019)]{Kooi2019}
Merel Kooi and Albert~A Koelmans.
\newblock {Simplifying Microplastic via Continuous Probability Distributions
  for Size, Shape, and Density}.
\newblock \emph{Environmental Science and Technology Letters}, 6:\penalty0
  551--557, 2019.
\newblock \doi{10.1021/acs.estlett.9b00379}.

\bibitem[Kooi et~al.(2017)Kooi, {Van Nes}, Scheffer, and Koelmans]{Kooi2017}
Merel Kooi, Egbert~H. {Van Nes}, Marten Scheffer, and Albert~A. Koelmans.
\newblock {Ups and Downs in the Ocean: Effects of Biofouling on Vertical
  Transport of Microplastics}.
\newblock \emph{Environmental Science and Technology}, 51\penalty0
  (14):\penalty0 7963--7971, 2017.
\newblock ISSN 15205851.
\newblock \doi{10.1021/acs.est.6b04702}.

\bibitem[Kukulka et~al.(2012)Kukulka, Proskurowski, Mor{\'{e}}t-Ferguson,
  Meyer, and Law]{Kukulka2012}
T.~Kukulka, G.~Proskurowski, S.~Mor{\'{e}}t-Ferguson, D.~W. Meyer, and K.~L.
  Law.
\newblock {The effect of wind mixing on the vertical distribution of buoyant
  plastic debris}.
\newblock \emph{Geophysical Research Letters}, 39\penalty0 (7):\penalty0 1--6,
  2012.
\newblock ISSN 00948276.
\newblock \doi{10.1029/2012GL051116}.

\bibitem[Lebreton et~al.(2019)Lebreton, Egger, and Slat]{Lebreton2019}
Laurent Lebreton, Matthias Egger, and Boyan Slat.
\newblock {A global mass budget for positively buoyant macroplastic debris in
  the ocean}.
\newblock \emph{Scientific Reports}, 9:\penalty0 12922, 2019.
\newblock ISSN 2045-2322.
\newblock \doi{10.1038/s41598-019-49413-5}.
\newblock URL \url{http://dx.doi.org/10.1038/s41598-019-49413-5}.

\bibitem[Liubartseva et~al.(2018)Liubartseva, Coppini, Lecci, and
  Clementi]{Liubartseva2018}
S~Liubartseva, G~Coppini, R~Lecci, and E~Clementi.
\newblock {Tracking plastics in the Mediterranean : 2D Lagrangian model}.
\newblock \emph{Marine Pollution Bulletin}, 129\penalty0 (February):\penalty0
  151--162, 2018.
\newblock \doi{10.1016/j.marpolbul.2018.02.019}.

\bibitem[Menna et~al.(2017)Menna, Gerin, Bussani, and Poulain]{Menna2017}
Milena Menna, Riccardo Gerin, Antonio Bussani, and Pierre-marie Poulain.
\newblock {The OGS Mediterranean Drifter Dataset: 1986-2016}.
\newblock Technical report, 2017.

\bibitem[Newman(2005)]{Newman2005}
M.~E.J. Newman.
\newblock {Power laws, Pareto distributions and Zipf's law}.
\newblock \emph{Contemporary Physics}, 46\penalty0 (5):\penalty0 323--351,
  2005.
\newblock ISSN 00107514.
\newblock \doi{10.1080/00107510500052444}.

\bibitem[Onink et~al.(2019)Onink, Wichmann, Delandmeter, and van
  Sebille]{Onink2019}
Victor Onink, David Wichmann, Philippe Delandmeter, and Erik van Sebille.
\newblock {The Role of Ekman Currents, Geostrophy, and Stokes Drift in the
  Accumulation of Floating Microplastic}.
\newblock \emph{Journal of Geophysical Research: Oceans}, 124\penalty0
  (3):\penalty0 1474--1490, 2019.
\newblock ISSN 21699291.
\newblock \doi{10.1029/2018JC014547}.
\newblock URL \url{https://doi.org/10.1029/2018JC014547}.

\bibitem[Pedrotti et~al.(2016)Pedrotti, Petit, Elineau, Bruzaud, Crebassa,
  Dumontet, Mart{\'{i}}, Gorsky, and C{\'{o}}zar]{Pedrotti2016}
Maria~Luiza Pedrotti, St{\'{e}}phanie Petit, Amanda Elineau, St{\'{e}}phane
  Bruzaud, Jean~Claude Crebassa, Bruno Dumontet, Elisa Mart{\'{i}}, Gabriel
  Gorsky, and Andr{\'{e}}s C{\'{o}}zar.
\newblock {Changes in the floating plastic pollution of the mediterranean sea
  in relation to the distance to land}.
\newblock \emph{PLoS ONE}, 11:\penalty0 1--14, 2016.
\newblock ISSN 19326203.
\newblock \doi{10.1371/journal.pone.0161581}.

\bibitem[Poulain et~al.(2019)Poulain, Mercier, Brach, Martignac, Routaboul,
  Perez, Desjean, and Halle]{Poulain2019}
Marie Poulain, Matthieu~J Mercier, Laurent Brach, Marion Martignac, Corinne
  Routaboul, Emile Perez, Marie~Christine Desjean, and Alexandra Halle.
\newblock {Small Microplastics As a Main Contributor to Plastic Mass Balance in
  the North Atlantic Subtropical Gyre}.
\newblock \emph{Environmental Science {\&} Technology}, 53:\penalty0
  1157--1164, 2019.
\newblock ISSN 0013-936X.
\newblock \doi{10.1021/acs.est.8b05458}.

\bibitem[Reisser et~al.(2015)Reisser, Slat, Noble, {Du Plessis}, Epp, Proietti,
  {De Sonneville}, Becker, and Pattiaratchi]{Reisser2015}
J.~Reisser, B.~Slat, K.~Noble, K.~{Du Plessis}, M.~Epp, M.~Proietti, J.~{De
  Sonneville}, T.~Becker, and C.~Pattiaratchi.
\newblock {The vertical distribution of buoyant plastics at sea: An
  observational study in the North Atlantic Gyre}.
\newblock \emph{Biogeosciences}, 12\penalty0 (4):\penalty0 1249--1256, 2015.
\newblock ISSN 17264189.
\newblock \doi{10.5194/bg-12-1249-2015}.

\bibitem[Ruiz-Orej{\'{o}}n et~al.(2018)Ruiz-Orej{\'{o}}n, Sard{\'{a}}, and
  Ramis-Pujol]{Ruiz-Orejon2018}
Luis~F. Ruiz-Orej{\'{o}}n, Rafael Sard{\'{a}}, and Juan Ramis-Pujol.
\newblock {Now, you see me: High concentrations of floating plastic debris in
  the coastal waters of the Balearic Islands (Spain)}.
\newblock \emph{Marine Pollution Bulletin}, 133\penalty0 (June):\penalty0
  636--646, 2018.
\newblock ISSN 18793363.
\newblock \doi{10.1016/j.marpolbul.2018.06.010}.
\newblock URL \url{https://doi.org/10.1016/j.marpolbul.2018.06.010}.

\bibitem[Ryan(2015)]{Ryan2015}
Peter~G. Ryan.
\newblock {Does size and buoyancy affect the long-distance transport of
  floating debris?}
\newblock \emph{Environmental Research Letters}, 10\penalty0 (8):\penalty0
  84019, 2015.
\newblock ISSN 17489326.
\newblock \doi{10.1088/1748-9326/10/8/084019}.
\newblock URL \url{http://dx.doi.org/10.1088/1748-9326/10/8/084019}.

\bibitem[Song et~al.(2017)Song, Hong, Jang, Han, Jung, and Shim]{Song2017}
Young~Kyoung Song, Sang~Hee Hong, Mi~Jang, Gi~Myung Han, Seung~Won Jung, and
  Won~Joon Shim.
\newblock {Combined Effects of UV Exposure Duration and Mechanical Abrasion on
  Microplastic Fragmentation by Polymer Type}.
\newblock \emph{Environmental Science and Technology}, 51\penalty0
  (8):\penalty0 4368--4376, 2017.
\newblock ISSN 15205851.
\newblock \doi{10.1021/acs.est.6b06155}.

\bibitem[Stokes(1847)]{Stokes1847}
G.~Stokes.
\newblock {On the theory of oscillatory waves}.
\newblock \emph{Transactions of the Cambridge Philosophical Society},
  8\penalty0 (441):\penalty0 197--229, 1847.
\newblock \doi{10.1057/9781137341280.0042}.

\bibitem[Suaria et~al.(2016)Suaria, Avio, Mineo, Lattin, Magaldi, Belmonte,
  Moore, Regoli, and Aliani]{Suaria2016}
Giuseppe Suaria, Carlo~G. Avio, Annabella Mineo, Gwendolyn~L. Lattin,
  Marcello~G. Magaldi, Genuario Belmonte, Charles~J. Moore, Francesco Regoli,
  and Stefano Aliani.
\newblock {The Mediterranean Plastic Soup: Synthetic polymers in Mediterranean
  surface waters}.
\newblock \emph{Scientific Reports}, 6:\penalty0 1--10, 2016.
\newblock ISSN 20452322.
\newblock \doi{10.1038/srep37551}.
\newblock URL \url{http://dx.doi.org/10.1038/srep37551}.

\bibitem[Turcotte(1986)]{Turcotte1986}
D.~L. Turcotte.
\newblock {Nr exp}.
\newblock \emph{Journal of Geophysical Research}, 91\penalty0 (B2):\penalty0
  1921--1926, 1986.

\bibitem[van Sebille et~al.(2020)van Sebille, Aliani, Law, Maximenko, Alsina,
  Bagaev, Bergmann, Chapron, Chubarenko, and C{\'{o}}zar]{Sebille2020}
Erik van Sebille, Stefano Aliani, Kara~Lavender Law, Nikolai Maximenko,
  Jos{\'{e}}~M Alsina, Andrei Bagaev, Melanie Bergmann, Bertrand Chapron, Irina
  Chubarenko, and Andr{\'{e}}s C{\'{o}}zar.
\newblock {The physical oceanography of the transport of floating marine
  debris}.
\newblock \emph{Environmental Research Letters}, 15\penalty0 (2):\penalty0
  023003, 2020.
\newblock ISSN 1748-9326.
\newblock \doi{10.1088/1748-9326/ab6d7d}.
\newblock URL \url{http://dx.doi.org/10.1088/1748-9326/ab6d7d}.

\bibitem[Virkar and Clauset(2014)]{Virkar2014}
Yogesh Virkar and Aaron Clauset.
\newblock {Power-law distributions in binned empirical data}.
\newblock \emph{Annals of Applied Statistics}, 8\penalty0 (1):\penalty0
  89--119, 2014.
\newblock ISSN 19417330.
\newblock \doi{10.1214/13-AOAS710}.

\end{thebibliography}

\end{document}